%% file: main.tex
\documentclass[sigconf]{acmart}

\usepackage{color, pifont, comment}
\usepackage{xspace}
\usepackage{xcolor}
\usepackage[ligature, inference]{semantic}
\usepackage{amsmath}
\usepackage{graphicx}
\usepackage{listings}
\usepackage{color, colortbl}
\usepackage{textcomp}
\usepackage{multirow}
\usepackage{makecell}
\usepackage{hyperref}
\usepackage[ruled, vlined, linesnumbered]{algorithm2e}
\usepackage[all]{xy}
\usepackage{tabularx}
\usepackage{caption}
\usepackage{subcaption}
\usepackage{booktabs}
\usepackage{tikz}

\AtBeginDocument{%
  \providecommand\BibTeX{{%
    \normalfont B\kern-0.5em{\scshape i\kern-0.25em b}\kern-0.8em\TeX}}}

\setcopyright{none}
\copyrightyear{}
\acmYear{}
\acmDOI{}

\acmConference[PrivaCI '22]{ACM Conference}{September 2022}{NY, USA}
\acmBooktitle{}
\acmPrice{}
\acmISBN{}

\newcommand{\ie}{{\it i.e.,}\xspace}
\newcommand{\eg}{{\it e.g.,}\xspace}
\newcommand{\etc}{{\it etc.}\xspace}

\newcommand{\etal}{{\it et al.}\xspace}
\newcommand{\vs}{{\it vs.}\xspace}
\newcommand{\wrt}{{\it w.r.t.}\xspace}

\newcommand{\policheckdataflow}{\textit{$\langle$data type, entity$\rangle$}}

\newcommand{\polichecktuple}{\textit{$\langle$app, data type, entity$\rangle$}}

\newcommand{\policheck}{PoliCheck}
\newcommand{\polisys}{Polisis}
\newcommand{\policylint}{PolicyLint}

\definecolor{maroon}{cmyk}{0,0.87,0.68,0.32}
\definecolor{brandeisblue}{rgb}{0.0, 0.44, 1.0}

\newcommand{\mycomment}[1]{}

\newcommand{\squishcount}{
   \begin{list}{\arabic{enumi})}
     { \usecounter{enumi}
       \setlength{\itemindent}{0em}
       \setlength{\parskip}{0pt}
      \setlength{\itemsep}{0pt}      \setlength{\parsep}{1pt}
      \setlength{\topsep}{1pt}       \setlength{\partopsep}{0pt}
      \setlength{\leftmargin}{1.5em} \setlength{\labelwidth}{10em}
      \setlength{\labelsep}{0.5em} } }
\newcommand{\countend}{
  \end{list}}

\begin{document}

\title{A CI-based Auditing Framework for Data Collection Practices}

\author{Athina Markopoulou}
\affiliation{%
 \institution{University of California, Irvine \\ athina@uci.edu}
 \country{USA}
}

\author{Rahmadi Trimananda}
\affiliation{%
  \institution{University of California, Irvine \\ rtrimana@uci.edu}
  \country{USA}
}

\author{Hao Cui}
\affiliation{%
  \institution{University of California, Irvine \\ cuih7@uci.edu}
  \country{USA}
}

\renewcommand{\shortauthors}{Markopoulou et al.}





\input{abstract}

\maketitle

\input{introduction}

\input{auditing}

\input{future}

\bibliographystyle{plain}
\bibliography{master,online}

\end{document}

%% file: abstract.tex
\begin{abstract}
Apps and devices (mobile devices, web browsers, IoT, VR, voice assistants, \etc{}) routinely collect user data, and send them to first- and third-party servers through the network. Recently, there is a lot of interest in (1) auditing the actual data collection practices of those systems; and also in (2) checking the consistency of those practices against the statements made in the corresponding privacy policies. In this paper\footnote{This position paper is supported by NSF Award 1956393. The paper was first presented at the 4th Annual Symposium on Applications of Contextual Integrity, NYC, Sept. 2022.}, we argue that the contextual integrity (CI) tuple can be the basic building block for defining and implementing such an auditing framework. We elaborate on the special case where the tuple is partially extracted from the network traffic generated by the end-device of interest, and partially from the corresponding privacy policies using natural language processing (NLP) techniques. Along the way, we discuss related bodies of work and representative examples that fit into that framework. More generally, we believe that CI can be the building block not only for auditing at the edge, but also for specifying privacy policies and system APIs.  We also discuss limitations and directions for future work.
\end{abstract}

%% file: introduction.tex
\section{The Problem Space}
\label{introduction}

 Personal data are routinely collected on end devices (browsers, mobile and IoT devices, smart TVs, VR devices, \etc{}) and shared with many first- and third-party entities,
without providing much transparency or control to users.  
Increased public awareness has led to data protection legislation, such as the GDPR,  
CCPA/CPRA, and other state or sector-specific data protection laws. 
These laws state rights of consumer or citizens, and duties of entities that collect, share, and use personal data.  Government agencies, such as the U.S. Federal Trade Commission (FTC), take initiatives to enforce those regulations. 
Their efforts are amplified by non-profits, privacy-advocates, and academics who report the results of their investigation on data collection practices and violations.
These developments have pushed tech companies, which were previously lacking incentives to self-regulate, towards the right direction, \ie to become increasingly  more transparent about their data collection practices and apply other good practices as well.

\textbf{The Gap between Systems and Laws.}
However, there is still a significant gap between the practices of tech companies, and the formulation and enforcement of privacy laws. First, from a software developer's perspective, it is often difficult to ensure compliance with all laws and regulations due to the complexity of their own system as well as their dependence on third party libraries, platforms, and other parts of the ecosystem that they have no control over. 
Second, policymakers need technical input to write relevant and enforceable policies. Sometimes the requirements in the laws do not directly map to a system specification that can be implemented and audited. Furthermore, the technology itself evolves rapidly and often renders privacy laws obsolete. 
The general problem of co-designing privacy-preserving systems along with consistent privacy laws---let alone auditing tools---is already quite daunting. Progress is made on narrow notions such as the meaning of ``singling out'' a user~\cite{cohen2020towards}, and this is an area of increasing research interest.

\begin{figure}[t!]
	\centering
	\includegraphics[width=\linewidth]{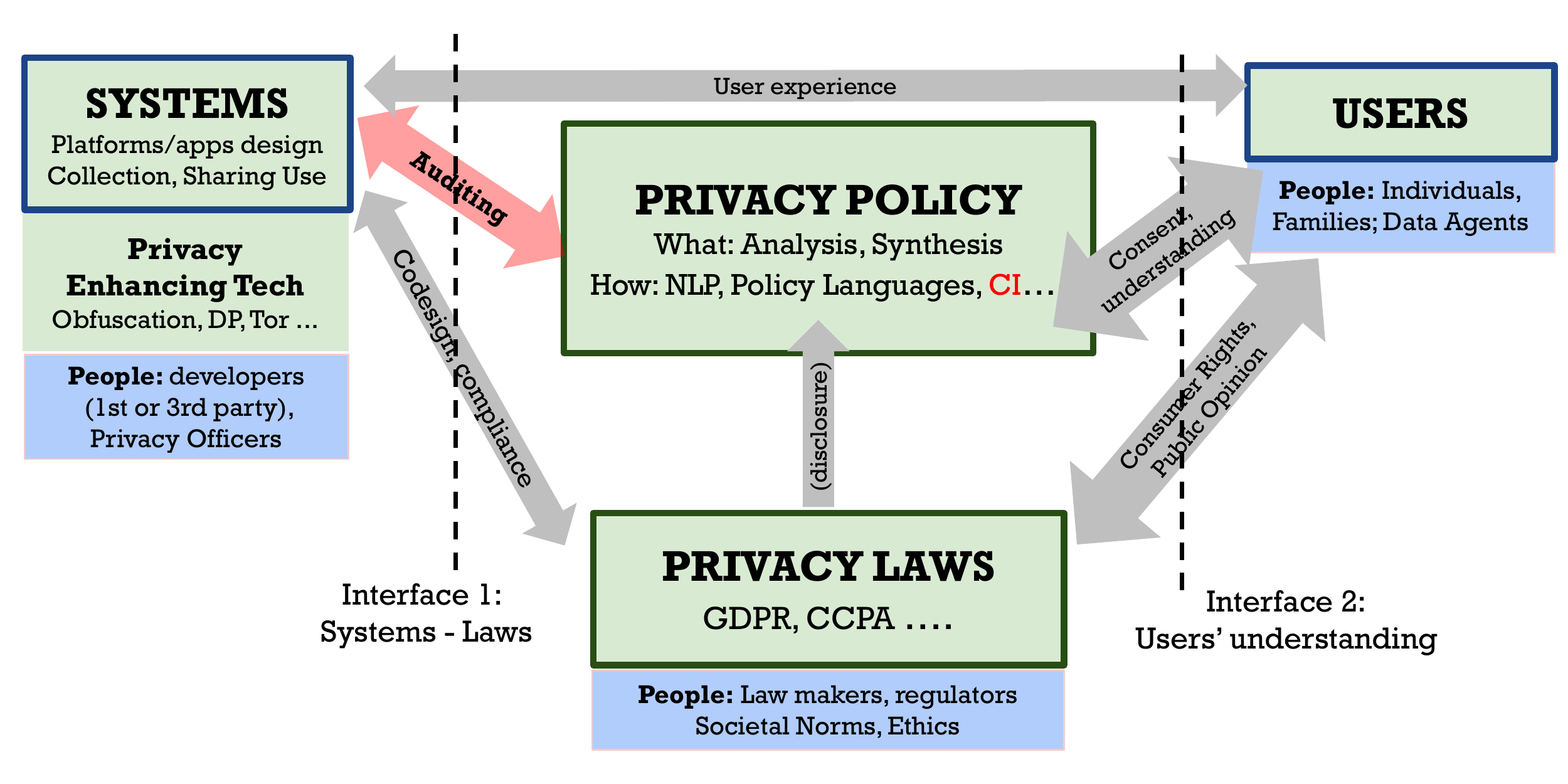}
	\caption{The privacy problem space involves computer systems, privacy laws, and users. An important component interfacing with all three is the privacy policy of the system---we presented an earlier version of this diagram in \cite{nsf-satc-breakout}. In this paper, we focus on the part highlighted in red: auditing the actual data collection practices of the system and the statements made in the corresponding privacy policy.}
	\label{fig:system-policy-law-users}
	\vspace{-10pt}
\end{figure}

\textbf{Transparency and Privacy Policies.} Although the general problem of technology-law interface is quite complex and difficult to tackle, there have been efforts and progress on developing methodologies to solve a more narrow problem: \textit{transparency}.
At the very least, most privacy laws require  that entities disclose their data collection, sharing, and use practices.  
For example, entities must disclose {\em what they collect},  {\em with whom they share it}, {\em for what purposes}, \etc{} Furthermore, most privacy laws also require that the user is notified and consents to those practices.
Software systems and services typically fulfill their ``notice-and-consent'' obligation through a {\em privacy policy} document that they provide to inform their users about their practices. Then, the user can choose to opt in or opt out before using the product or service.
Privacy policy documents are legally binding and they are the focus of this paper.
\footnote{There may be additional disclosures and privacy notices via blogs, press releases, websites, \etc{}, which further communicate to the public how an organization collects and processes personal data, and what rights it provides.}  
 An organization's written privacy policy is a central piece of the puzzle, as depicted on Figure~\ref{fig:system-policy-law-users}:
 
\squishcount
\item A privacy policy must be written so that it meets the minimum requirements  in order to be compliant with the privacy laws.
\item A privacy policy must  accurately and comprehensively describe the data collection practices implemented by the system and any third-party libraries the system uses.  
\item A privacy policy should be (i) comprehensive in its coverage of the system functionality and law requirements,  (ii) self-consistent, \ie without contradicting statements, and   present information to the user in a way that is easily understood.
\countend

\begin{figure}[t!]
	\centering
	\includegraphics[width=\linewidth]{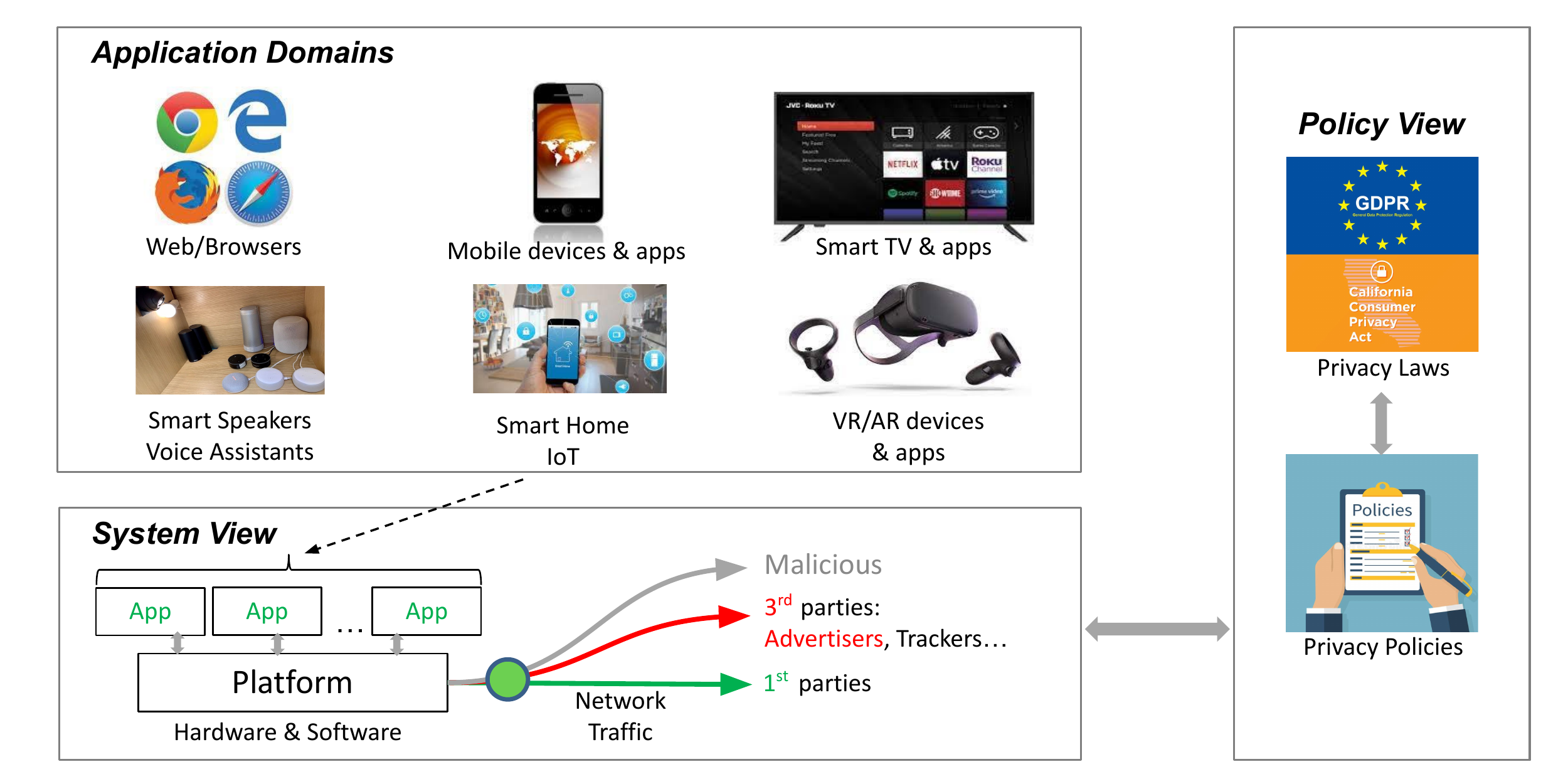}
	\caption{Data collection at the edge by apps, devices, and platforms. The outgoing network traffic provides a unique vantage point for monitoring the actual information flow.}
	\label{fig:app-system-law}
	\vspace{-10pt}
\end{figure}

\textbf{Auditing Data Collection at the Edge.} To further narrow down the scope of the problem, let us consider Figure \ref{fig:app-system-law}. It depicts various end-devices and platforms that users interact with, including mobile devices, browsers, smart TVs, VR devices, IoT devices, \etc{}. These are different platforms that enable their respective apps. In all these app ecosystems,  personal data are collected on devices (from the apps, platforms, and third-party libraries), and are sent over the Internet to first- and third-party entities for functional, advertising and tracking services (ATS), and many other purposes.
%
The research community has followed different types of broad approaches for auditing data collection practices at the edge: 
\squishcount
\item A large body of work obtains and analyzes the {\em actual} information flow observed into and out of an app, device, or platform using a range of techniques, \ie{} static~\cite{androidappstudy} and dynamic analysis~\cite{taintdroid}, and network traffic analysis~\cite{mohajeri2019watching,varmarken2020tv,recon,razaghpanah2018apps,shuba2016antmonitor,shuba2020nomoats,shuba2018nomoads}.
\item Another large body of work  extracts the {\em intended/declared} information flow. This is mostly done by analyzing the privacy policy of the system; it  originally relied on experts reading the policies, but is getting increasingly automated using NLP~\cite{harkous2018polisis,andow2019policylint}.  Another source for extracting the {\em intended/declared} information flow is through permissions~\cite{lentzsch2021heyalexa}.
\item There is also a third body of work that checks the {\em consistency} between the intended/declared information flow in the privacy policy, and the actual behavior found in the system (\eg in the network traffic~\cite{andow2020actions}).
\countend

Our key observation is that in all the above cases, the extracted information is {\em part} of the parameters of the CI-tuple:\footnote{When analyzing information flow at the edge, the ``subject'' in the CI-tuple is the user of the device/app, thus will be omitted for lack of space in this paper.}

\noindent\fbox{$\langle sender, recipient, data~type, [subject], transmission~principle\rangle$}
\vspace{5pt}

By observing the actual behavior of the system or its declared behavior in its privacy policy, one can observe and automatically extract meaningful values from each CI parameter as summarized in Figure~\ref{fig:ci-tuple}.
The three aforementioned bodies of work typically obtain and report part of the tuple, often without recognizing that it fits into the CI framework~\cite{privacy-in-context}.
 
\textbf{Position.}  We believe that the CI-tuple is well-suited to be the central building block (data structure) for specifying and auditing data collection practices at the edge, together with their privacy policies. Not only that it is intuitive and interpretable, but it also lends itself to automation and large-scale auditing. Representative examples are discussed in Section~\ref{sec:ci-based-auditing}.  Limitations and possible extensions  are discussed in Section~\ref{sec:open-problems}. 


%% file: auditing.tex
\section{CI-Based Auditing}
\label{sec:ci-based-auditing}


\subsection{Network Traffic Monitoring} 

\begin{figure}[t!]
	\centering
	\includegraphics[width=\linewidth]{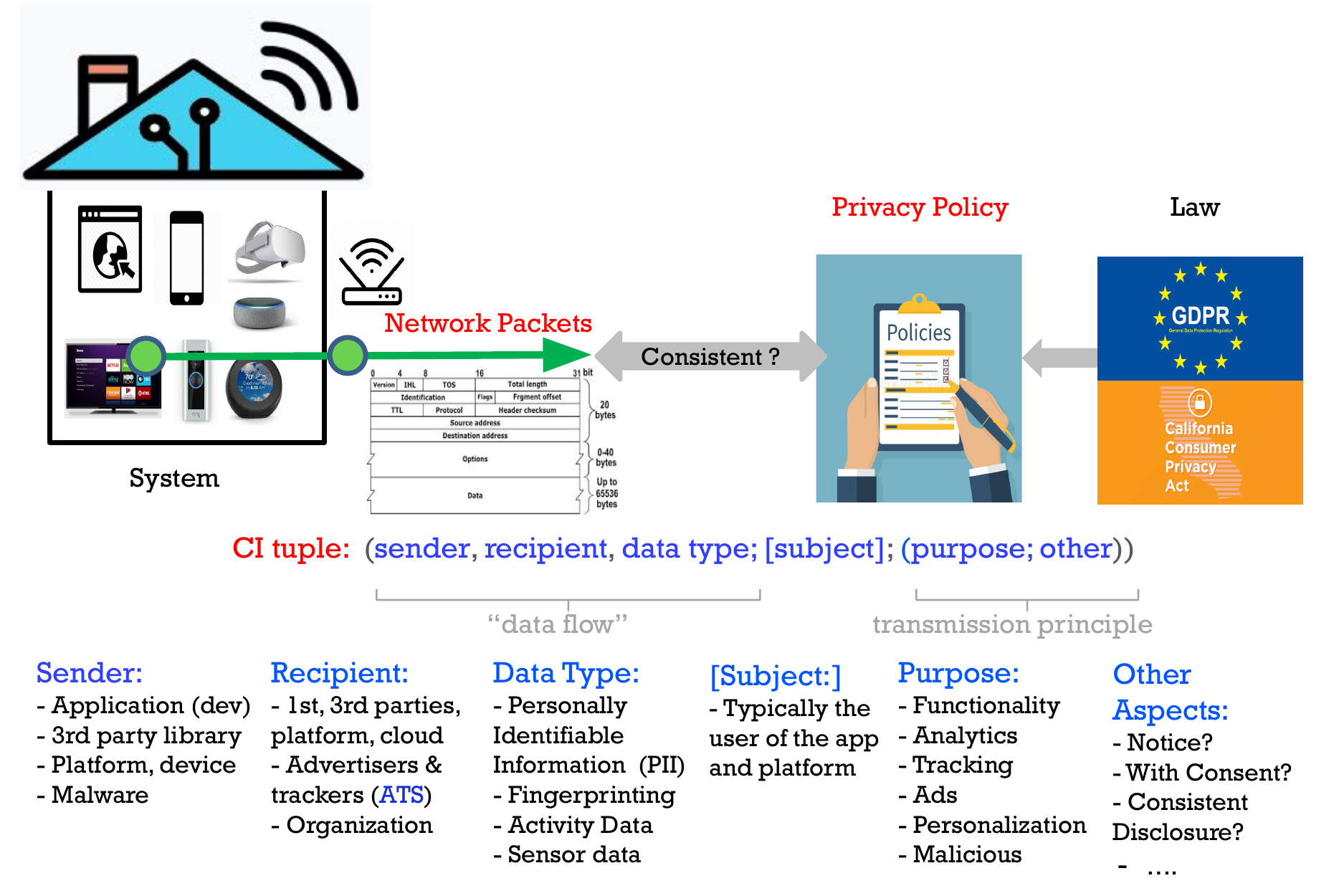}
	\caption{Extracting (parts of) the CI tuple from the network traffic and the privacy policy. 
	}
	\label{fig:ci-tuple}
	\vspace{-10pt}
\end{figure}

Apps, platforms, and third-party libraries running on a device can collect personal data and send them to remotes server over the Internet, as depicted in Figures~\ref{fig:app-system-law} and~\ref{fig:ci-tuple}. Monitoring the network traffic coming out of the device provides a unique vantage point to observe data collection: \textit{who} collects \textit{what}, and \textit{where} it is sent.

Intercepting and inspecting network traffic, can be done on the device itself (examples on Android include our own AntMonitor~\cite{shuba2016antmonitor} and Lumen~\cite{razaghpanah2018apps}), or using a proxy~\cite{mitmproxy}. Packet traces are collected, and typically stored and analyzed offline. By observing parts of the packets that are sent in the clear, it is trivial to extract the {\em recipient} of the information flow, \eg{} through destination domains (DNS queries), IP addresses, reverse DNS lookups, \etc{} It is also trivial to extract further information about the organization/entity a destination domain belongs to (\eg{} using ``whois'' or other lookup methods), and classify the domain as first- (\eg{} if the domain matches or partially contains the name of the entity), or third-party (\eg{} it is categorized as an ATS domain based on well-known blocklists). This labeling as first- vs. third-party also indicates the {\em purpose} of data collection. Furthermore, when network traffic collection is done on the device~\cite{shuba2016antmonitor,razaghpanah2018apps}, it is also trivial to extract the {\em sender} (\ie{} app, platform, or third-party library). 
%
It is more difficult to extract the {\em data types} sent in network packets, since the packets are typically encrypted. While this encryption can be bypassed, it is oftentimes not clear what to look for in the payload of a decrypted packet; the payload usually contains key-value pairs, whose meaning often cannot be interpreted precisely. In summary, from collected network packet traces at the edge, we can extract several parameters of the CI-tuple:

\noindent
\fbox{$F'=\langle sender=platform/app, destination, data~type, purpose \rangle$}


Several researchers in the Internet measurement community have performed network traffic analysis on different platforms to characterize their data collection and sharing practices focusing on ATS. For example, just within our group, we have systematically tested the apps and inspected the traffic generated by mobile devices and their apps~\cite{shuba2018nomoads, shuba2020nomoats}, smart TV platforms and their apps~\cite{varmarken2020tv, varmarken2022fingerprintv}, IoT devices~\cite{trimananda2020packet}, VR headsets~\cite{trimananda2022ovrseen}, and most recently smart speakers~\cite{iqbal2022your}. We have also characterized and reported the aforementioned CI parameters. Many other groups have done similar studies for these platforms; examples include ~\cite{razaghpanah2018apps,apthorpe2018keeping,jin2018mobipurpose,mohajeri2019watching,recon}, but the list is by no means exhaustive---this is an active field.




\subsection{Privacy Policy Analysis}
Privacy policy and consistency analysis is becoming increasingly automated using NLP, and applied across various app ecosystems ~\cite{zimmeck2014privee,slavin2016toward,zimmeck2017automated,wang2018guileak,harkous2018polisis,andow2019policylint, andow2020actions}; this is also a large body of work to properly review here.
Instead, we describe state-of-the-art tools, namely \policylint{}~\cite{andow2019policylint}, \policheck{}~\cite{andow2020actions}, and \polisys{}~\cite{harkous2018polisis} that demonstrate what is possible w.r.t. CI. 
They all apply NLP on: (1) the text of privacy policies to extract and analyze data collection and sharing statements; and (2) compare them against the actual data flows found in network traffic (\textit{flow-to-policy} consistency analysis). 

 \policylint{}~\cite{andow2019policylint} provides an NLP pipeline that takes a sentence as input. For example, it takes {\em ``We may collect your email address and share it for advertising purposes''}, and extracts the collection statement ``(entity: \emph{we}, action: \emph{collect}, data type: \emph{email address})''.
More generally, \policylint{} takes the app's privacy policy text, parses sentences, performs NLP techniques, and eventually extracts data collection statements defined as the tuple $P=$\polichecktuple{};
\textit{app} is the sender and \textit{entity} is the recipient organization/entity performing an \textit{action} (``collect'' or ``not collect'') on the \textit{data type}, 
and outputs:

 \centerline{\fbox{$P=\langle sender=platform/app, recipient=entity, data~type\rangle$}}
 

\policheck{}~\cite{andow2020actions}  takes the app's ``data flows'' (extracted from the network traffic and defined as $F=$\policheckdataflow{} and compares it against the stated $P$ for consistency. 
\policheck{} classifies the disclosure of $F$ as {\em clear} (if the data flow exactly matches a collection statement), {\em vague} (if the data flow  matches a collection statement in broader terms), {\em omitted} (if there is no collection statement corresponding to the data flow), {\em ambiguous} (if there are contradicting collection statements about a data flow), or {\em incorrect} (if there is a data flow for which the collection statement states otherwise).\footnote{Following \policheck{}'s terminology~\cite{andow2020actions},  these five types of disclosures can be grouped into  two groups: \emph{consistent} (clear and vague disclosures) and \emph{inconsistent} (omitted, ambiguous, and incorrect) disclosures. 
For consistent disclosures,  there is a statement in the policy that matches the data type and entity, either clearly or vaguely.}

\begin{figure}[t!]
    \centering
    \begin{subfigure}[b]{0.5\textwidth}
        \centering
        \includegraphics[width=\textwidth]{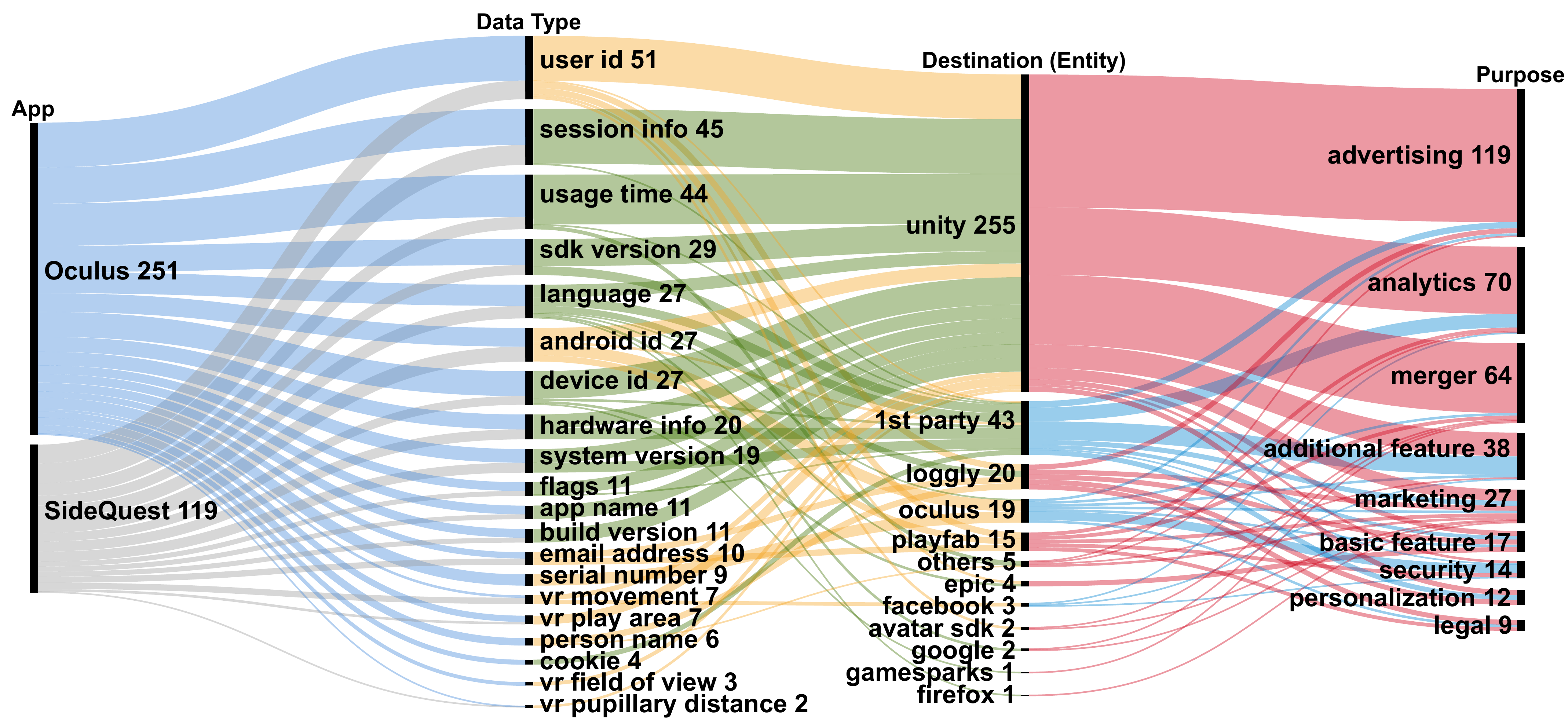}
        \caption{$P'$ tuples: augmenting $P$ (from \policheck{}) with {\em purpose} (from \polisys{}). }
        \label{fig:tuple-with-purpose}
    \end{subfigure}
    \hfill
    \begin{subfigure}[b]{0.4\textwidth}
        \centering
        \includegraphics[width=\textwidth]{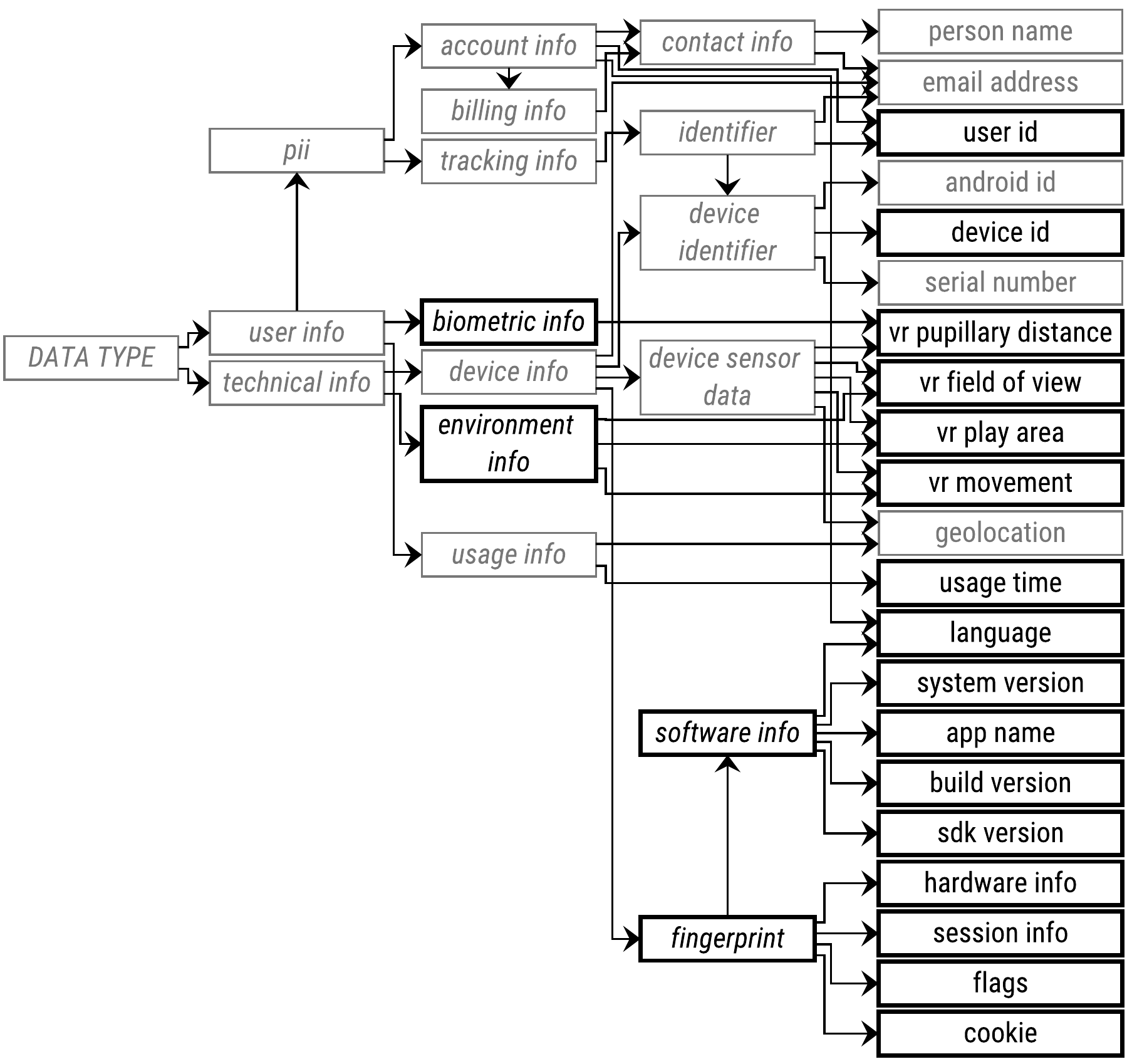}
        \caption{VR-specific data ontology}
        \label{fig:entity-ontology}
    \end{subfigure}
    \vspace{-8pt}
    \caption{Case study: auditing network traffic and privacy policies of the most popular Oculus VR apps; see  \cite{trimananda2022ovrseen}.}
    \vspace{-1em}
    \label{fig:tuple-and-ontology}
\end{figure}

Consistency analysis  between the ``data flows'' ($F$ or its augmented version $F'$) found in network traffic and collection statements ($P$ or its augmented version $P'$) rely on pre-built ontologies and synonym lists. These are used to match (i) the data type and destination that appear in each $F$ with (ii) any instance of $P$ that discloses the same (or a broader) data type and destination\footnote{For example see Figure~\ref{fig:tuple-and-ontology}(b): ``email address'' is a special case of ``contact info'' and, in turn, of ``pii''. There is a clear disclosure \wrt data type if the ``email address'' is found in a data flow and a collection statement. A vague disclosure is declared if the ``email address'' is found in a data flow and a collection statement that uses the term ``pii'' is in the privacy policy. An omitted disclosure means that ``email address'' is found in a data flow, but there is no mention of it (or any of its broader terms) in the privacy policy.}.

\policylint{} and \policheck{} were originally developed for mobile apps and later applied to Alexa skills \cite{lentzsch2021heyalexa} and Oculus VR apps \cite{trimananda2022ovrseen}. In  \cite{trimananda2022ovrseen},  we developed Oculus VR-specific ontologies; the  data ontology is shown on Figure~\ref{fig:tuple-and-ontology}(b). Furthermore, we improved several aspects of PoliCheck and augmented the $P$ tuple with {\em purpose}, which we were able to extract from the privacy policy using \polisys{}~\cite{harkous2018polisis}. The augmented tuples are visually depicted on Figure~\ref{fig:tuple-and-ontology}(a):

 \centerline{\fbox{$P'=\langle sender=platform/app, recipient=entity, data~type, purpose\rangle$}} 

{\em Purpose/use} is an important dimension of the transmission principle (TP) parameter in CI and there are automated ways to automatically extract purpose from privacy policies. Other state-of-the-art methods beyond  \polisys{} ~\cite{harkous2018polisis} include the following.  MobiPurpose~\cite{jin2018mobipurpose} inferred data collection purposes of mobile apps using network traffic and app features (\eg URL paths, app metadata, domain name, \etc). PurPliance~\cite{bui2021consistency} automates the inference of data collection purposes introduced in MobiPurpose, extracts purposes from the privacy policy, and performs flow-to-policy consistency analysis that includes checking the consistency of these purposes.

\subsection{Extracting the CI tuple}

A key observation is that existing methods for monitoring and reporting information flow from an end device already extract (a subset or all of) the CI parameters in the CI-tuple. However, they currently do it in an ad-hoc way, often without realizing that these methods fit into the CI framework, with a few exceptions ~\cite{shvartzshnaider2019going}. Furthermore, this extraction can be done through an {\em automated} pipeline: automatically testing the behavior of a large number of apps, collecting network traffic, extracting the ``data flows'' $P$ found in the traffic, performing NLP techniques to analyze privacy policies, and extracting the collection and sharing statements $F$. Extracting the well-defined parameters of the CI-tuple, \ie{}  $P'=\langle sender=platform/app, recipient=entity, data~type, subject=user \rangle$, is well understood by now. Even some values of the more elusive TP can be extracted automatically. For example in \cite{trimananda2022ovrseen}, we were able to extract the following dimensions of TP: (1) {\em consistency}, \ie{} whether actual data flows extracted from network traffic agree  with the corresponding statements made in the privacy policy; 
(2) {\em purpose}, extracted from privacy policies and confirmed by destination domains (\eg whether they are ATS); and 
(3) the presence of ``notice-and-consent'', while testing the apps. %

We believe that the CI-tuple can provide a unifying data structure, and interface for auditing both the intended/declared and actual end-device behaviors, as well as the consistency between the two. 

%% file: future.tex
\section{Open Problems \& Future Directions}
\label{sec:open-problems}

We have argued that parts for the CI-tuple is already used in practice for auditing systems and their policies, and that the CI-tuple should be proposed in a more intentional and unifying way as the data structure for auditing data collection and sharing practices. However, there are still open conceptual problems (beyond just defining TP) to address, before CI can realize this potential.

\begin{figure}[t!]
    \centering
    \begin{subfigure}[b]{0.40\textwidth}
        \centering
        \fbox{\includegraphics[width=\textwidth]{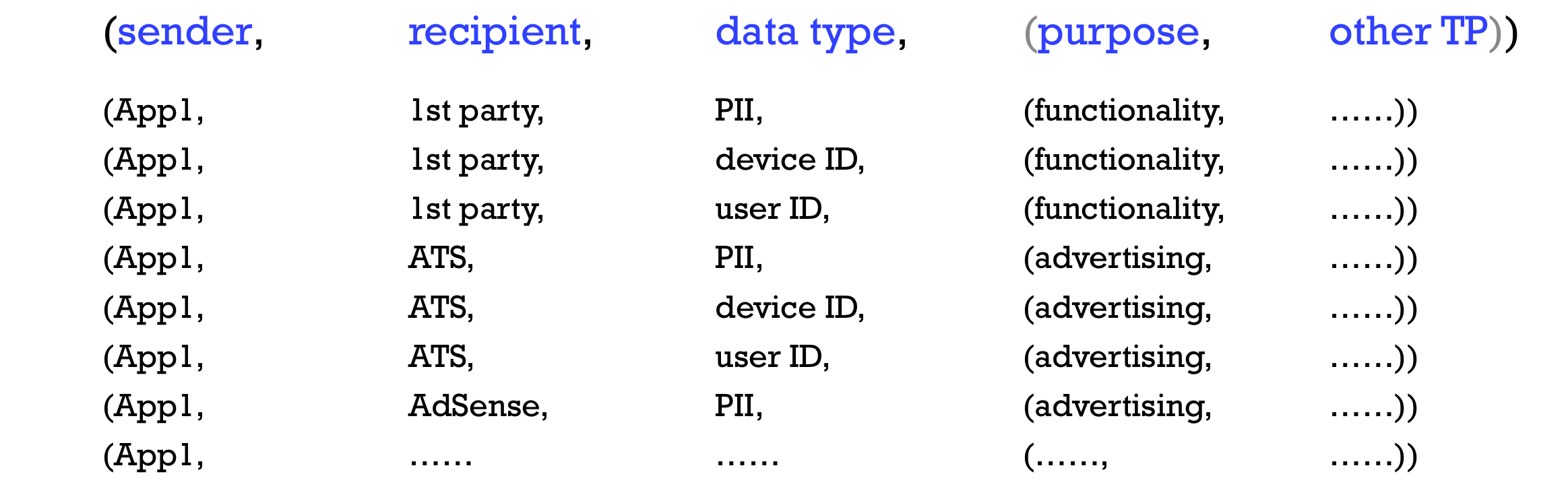}}
        \caption{``Flat'' CI tuples}
        \label{fig:tuple-flat}
    \end{subfigure}\\
    \vspace{10pt}
    \begin{subfigure}[b]{0.40\textwidth}
        \centering
        \fbox{\includegraphics[width=\textwidth]{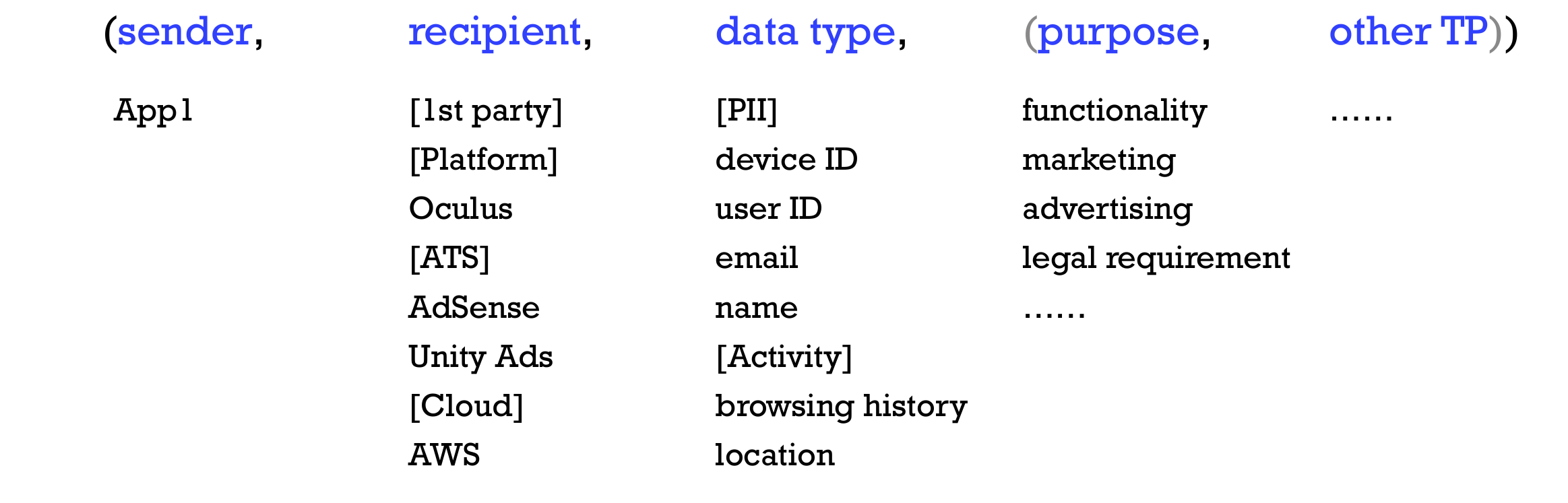}}
        \caption{``Bloated'' CI tuples}
        \label{fig:tuple-bloated}
    \end{subfigure}\\
    \vspace{10pt}
    \begin{subfigure}[b]{0.40\textwidth}
        \centering
        \fbox{\includegraphics[width=\textwidth]{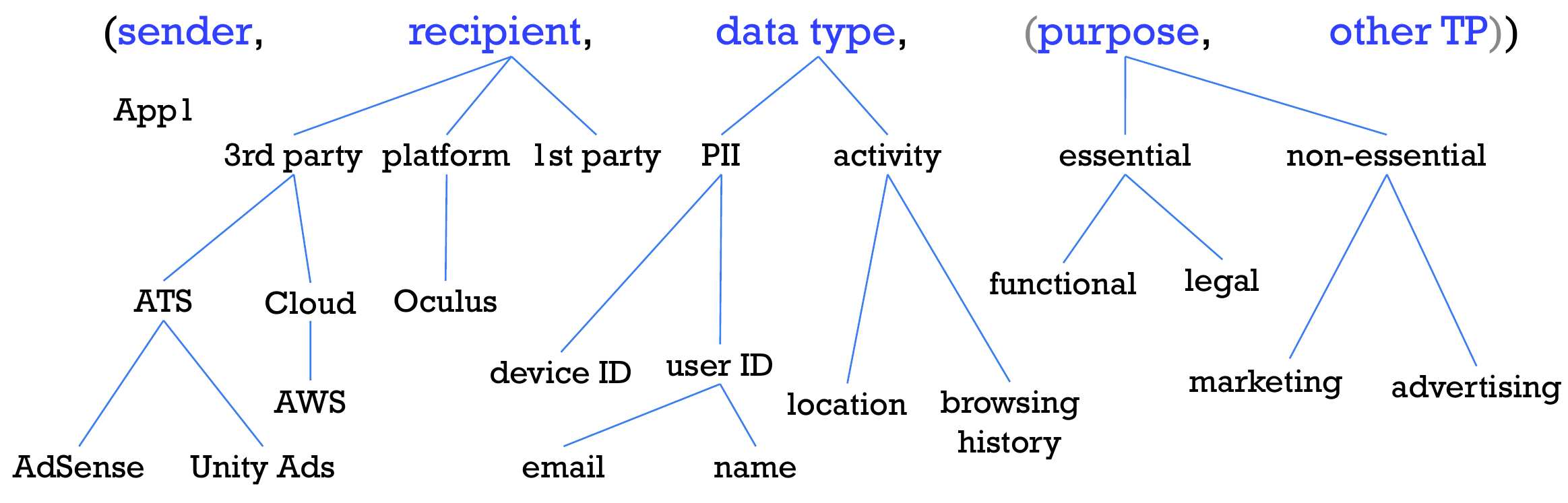}}
        \vspace{-10pt}
        \caption{``Hierarchical'' CI-tuples (\ie with ontologies).}
        \label{fig:tuple-tree}
    \end{subfigure}
    \caption{Open question: how to extract and summarize (a large number of) CI tuples? Note: the ``subject'' parameter is implicitly the user of the device or app, and it is omitted.}
    \label{fig:ci-tuples}
    \vspace{-10pt}
\end{figure}

\noindent \textbf{Q1: Dealing with hierarchical parameters.} The CI-tuple is by definition ``flat'', as depicted on Figure~\ref{fig:ci-tuples}(a). However, when we audit a device (both from the system and privacy policy perspectives), we collect multiple such tuples  capturing different information flows (one per packet or per collection statement, respectively). The values of the same CI-parameter can vary across different tuples. How should we summarize and process this information?

One option is to consider the values for each CI parameter as  unrelated, essentially a laundry list of possible values, as  in Figure~\ref{fig:ci-tuples}(b). 
Shvartzshnaider \etal  analyzed Facebook's privacy policy~\cite{shvartzshnaider2019going} and coined the term ``bloated'' policy. Considering the actual CI-tuple in its full context, as in Figure~\ref{fig:ci-tuples}(a), is a more precise disclosure. However, many privacy policies today (even the good ones like Unity's~\cite{unity-privacy-policy}) are organized in sections containing bulleted lists (\eg{} what is collected, with whom it is shared, for what purpose, \etc{}) to match what the law dictates and to also be more readable.  

There is one more complication: possible values of each CI parameter can be organized in an ontology: a conceptual example is shown on Figure~\ref{fig:ci-tuples}(c) and an example based on real data is shown on Figure~\ref{fig:tuple-and-ontology}(b). This hierarchy is not only intuitive, but also necessary for checking the consistency between a value observed in the network traffic and what is declared in the policy. Without an ontology, collecting an ``email'' would be inconsistent with the privacy policy that states the collection of ``PII'': the former is clearly subsumed by the latter. We believe that developing a concept of a ``hierarchical'' CI-tuple would be necessary for auditing.

\textbf{Q2: How to combine parameters extracted from different sources into a single CI-tuple.} Different sources (\eg{} network traffic, dynamic analysis, NLP on privacy policy, permissions, \etc{}) can reveal some or all the parameters of the CI-tuple. How does one merge that information from different sources into a single CI-tuple? If the value of the same CI parameter obtained from different sources is consistent, it can help link the corresponding tuples and add dimensions that were possibly missing.\footnote{This is what we did in the example in Figure~\ref{fig:tuple-and-ontology}(a): we took the flow-to-policy analysis result from \policheck{} that maps data flows $\langle destination, data~type \rangle$ from network traffic to policy statements, and mapped each statement to its corresponding policy segment. We then used \polisys{} to extract purposes from each segment, and reported the augmented tuple.} But how should one deal with inconsistencies? In~\cite{trimananda2022ovrseen}, we included consistency as part of the TP parameter, but there may be other ways worth exploring.

\textbf{Q3: Reactive vs. proactive use of CI.} What we  proposed so far is a {\em reactive/diagnostic} use of CI for auditing, namely to extract and report the tuple, so as to characterize the appropriateness of the information flow. In addition, we believe that CI can play a {\em proactive} role in defining the  specification of privacy policies and system APIs.  The sections currently found in privacy policies (see Unity's privacy policy~\cite{unity-privacy-policy} as an example) can follow the structure of the CI-tuple. Furthermore, platforms and their apps can be designed to provide APIs to return the CI-tuple for auditing purposes.

Finally, revisiting the bigger picture in Figure~\ref{fig:system-policy-law-users}, using the  CI-tuple to co-design privacy policies and systems can also be useful in other interactions, such as in the communication with users and in interfacing with privacy laws (possibly through the TP).